\documentclass{PoS}

\usepackage{graphicx}
\usepackage{epstopdf}
\usepackage{amssymb}
\usepackage{amsmath}
\usepackage{bbold}
\usepackage{psfrag}
\usepackage{epsfig}
\usepackage{cancel}

\newcommand{\bra}{\langle}
\newcommand{\ket}{\rangle}
\newcommand{\cH}{{\cal H}}

\newcommand{\x}{{\bf x}}
\newcommand{\xp}{{\bf x'}}
\newcommand{\y}{{\bf y}}
\newcommand{\p}{{\bf p}}
\newcommand{\kk}{{\bf k}}
\newcommand{\ad}{a^\dagger}
\newcommand{\kp}{{\bf k'}}

\newcommand{\pb}{\overline{P}}
\newcommand{\rb}{\overline{R}}
\newcommand{\rr}{{\bf r}}

\title{Volumes of Space as Subsystems}

\ShortTitle{Volumes of Space as Subsystems}

\author{\speaker{Federico Piazza}\\ 
        Perimeter Institute for Theoretical Physics\\
Waterloo, Ontario, N2L 2Y5, Canada}

\author{Fabio Costa\\
        Perimeter Institute for Theoretical Physics\\
Waterloo, Ontario, N2L 2Y5, Canada.\\
Dipartimento di Fisica, Universit\`a di Milano}

\abstract{As a novel approach with possible relevance to
semiclassical gravity, we propose to define regions of space
as quantum subsystems.
After recalling how to divide a generic quantum system into ``parts'', we apply this idea to a free scalar field in Minkowski space and we compare two different localization schemes. The first scheme is the standard one, induced by the local relativistic fields; the alternative scheme that we consider is the one induced by the Newton-Wigner operators. If degrees
of freedom are divided according to the latter, the Hamiltonian of the field exhibits a certain amount of non-locality. Moreover,
when a region of space is cut off from the rest according to the Newton-Wigner scheme,
the geometric entropy is finite and exhibits a sensible thermodynamic behaviour.}

\FullConference{From Quantum to Emergent Gravity: Theory and Phenomenology\\
		June 11-15 2007\\
		Trieste, Italy}

\begin{document}

\section{Subsystems and Local Subsystems}

Among other famously intriguing and counterintuitive aspects of quantum physics,
relatively little attention has been paid to the quantum mechanical description of a composite system.
The partitions of a quantum system -- i.e. all possible ways that a quantum system can 
be divided into ``parts'' -- have a mathematical structure completely different from, for example,
the partitions of a set in set theory. In set theory you can choose a bi-partition
A-B by going through each element of a (countable) set and deciding whether it belongs to subset A or B.
Clearly, finite sets admit only a finite number of possible partitions. Analogously, 
a finite lattice can be divided into sub-volumes in a finite number of ways. 

Quantum mechanics divides things differently \cite{paolo1}; a quantum system can be partitioned if its Hilbert space can be written as a \emph{tensor product} of Hilbert spaces. Consider, for instance, a system
described by the Hilbert space $\mathbb{C}^4$. The latter can be seen as a two-spin system and written as  $\mathbb{C}^4 = \mathbb{C}^2_A \otimes \mathbb{C}^2_B$, where indices $A$ and $B$ identify each of the two identical components $\mathbb{C}^2$. Given any orthonormal basis
$\{|a\ket,\, |b\ket,\, |c\ket,\, |d\ket\}$ of $\mathbb{C}^4$,  one way of partitioning the system is through the identification 
\begin{equation}\label{identification1}
|a\ket \simeq |0\ket_A \otimes |0\ket_B, \quad |b\ket \simeq |0\ket_A \otimes |1\ket_B, \quad |c\ket \simeq |1\ket_A \otimes |0\ket_B, \quad |d\ket \simeq |1\ket_A \otimes |1\ket_B, 
\end{equation}
where $\{|0\ket_{A}, \, |1\ket_{A}\}$ and $\{|0\ket_{B}, \, |1\ket_{B}\}$ are some choosen basis in $\mathbb{C}^2_{A}$ and $\mathbb{C}^2_{B}$ respectively. 
A different partition is defined by the choice of another orthonormal basis, say $\{|a'\ket,\, |b'\ket,\, |c'\ket,\, |d'\ket\}$, to use for the one to one correspondence \eqref{identification1}. All possible partitions of $\mathbb{C}^4$ are thus given by the elements of the group $SU(4)$ except that, within $SU(4)$, there 
are also transformations that merely correspond to a change of basis in either of the two factors $\mathbb{C}^2$. These transformations have to be factored out since they don't change the partition, leaving us with the group $SU(4)/SU(2)^2$: those are all the inequivalent ways we can separate a two-spin\footnote{More generally, a $d^N$-dimensional Hilbert space  can be partitioned into $N$ smaller systems each of dimension $d$ and such partitions are in one to one correspondence with the elements of  $SU(d^N)/SU(d)^N$} system! Note therefore that
quantum degrees of freedom, even when finite, can be split in an infinite number of ways.
Not only can you choose whether some of them belong to, say, subsystem $A$ or $B$, but, as opposed to the elements of a set or the sites of a lattice, you can unitarily mix them before the splitting, in such a way that they completely lose their individual identities. 

Many appealing arguments in semi-classical gravity, such as those related to black hole thermodynamics and to the holographic principle, are based upon the splitting of quantum degrees of freedom into two parts, each belonging to separate regions of space, typically across a causal horizon or just across some imaginary boundary.
According to the holographic principle \cite{hol}, when gravity is taken into account, the total number of degrees of freedom is bounded by the area -- rather than the volume -- of the region.  A breakdown of locality has  also been invoked (e.g. \cite{solu}) in relation  to the \emph{black hole information-loss paradox}. Such hints are clearly in conflict with local quantum field theory
and call for a deep reassessment of our current physical understanding.
Instead of venturing into the highly arbitrary and unknown realm of possible non-local theories, here we take the rather conservative point of view of maintaining the basic dynamics of our successful quantum theories (e.g. the Hamiltonian of the Standard Model) and just allowing some flexibility when it comes to dividing their quantum degrees of freedom according to distinct regions of space. 
Such a ``pre-geometric" approach \cite{fedo} may look a bit fictitious since, after all,
locality is already built into quantum field theory and the correct ``local'' 
tensor product structure should turn out to be the one actually induced by the local fields.
In the rest of this paper we question this established  point of view by comparing a few aspects of the standard localization procedure with those of an alternative one, induced by 
the ``Newton-Wigner'' \cite{nw} operators. Our main interest, rather than just philosophical, is to present a different 
rationale to be possibly applied in semiclassical gravity whenever it comes to isolating a bunch of ``local'' degrees of freedom.
A more thoroughgoing and operationally based case for alternative localization schemes will appear elsewhere \cite{costapiazza}.

Although the general approach that we are following here is fairly recent \cite{fedo},
Newton Wigner (NW) operators are almost 60 years old. In this paper we review aspects of the standard and NW 
localization schemes that, to some extent, are already known in the 
literature, but in the new light of \cite{fedo}. 
At the end of section \ref{sec:entropy} we also mention some new
results that will appear in more detail elsewhere \cite{anchesergio}.
We will work in the Schroedinger picture where observables do not evolve in time. We will be rather cavalier 
about the mathematical subtleties involved with continuous tensor products: 
say that IR and UV regulators are implicitly assumed which make 
the total dimension of $\cH$ finite. 

\section{The Two Localization Schemes}

It is possible to assign a tensor product structure (TPS) to a system by specifying a set of 
accessible observables \cite{zan2}. 
Consider a quantum system divided into two parts,
$P$ and $R$:  $\cH = \cH_P \otimes \cH_R$. $P$ stands for ``place'' and $R$ stands 
for ``rest of the system''. Which tensor decomposition actually divides $\cH$ into ``places'' 
is the matter of the present debate.
If we have two sets of observables, ${\cal A}^j_P$ and ${\cal A}^j_R$,  
separately defined in subsystem $P$ and $R$ respectively, then
we can trivially extend such observables to the entire system as follows,
\begin{equation}
{\cal A}^j_P\  \longrightarrow\ {\cal A}^j(P) \ \equiv \ {\cal A}^j_P \otimes \mathbb{1}_R, \quad \qquad
{\cal A}^j_R\  \longrightarrow\ {\cal A}^j(R) \ \equiv \ \mathbb{1}_P \otimes {\cal A}^j_R, 
\end{equation}
i.e. we just make them act as the identity on the other subsystem. By construction we have
\begin{equation}\label{commute}
[ {\cal A}^j(P), {\cal A}^k(R)]\, =\, 0.
\end{equation}
The basic idea here (see \cite{zan2} for more detail and mathematical rigor) 
is that the converse is also true.
That is if  we isolate two subalgebras ${\cal A}(P)$ and ${\cal A}(R)$, within the algebra of observables acting on $\cH$, satisfying \eqref{commute}, then they induce a unique\footnote{Actually, only if 
the two subalgebras generate the entire algebra of operators on $\cH$ \cite{zan2}}  
bipartition $\cH = \cH_P \otimes \cH_R$.
Since in quantum field theory (QFT) the usual local observables commute at space-like separated events, 
we have a straightforward realization of \eqref{commute} and we can use local fields to define 
a local TPS at each time $t$. 

At the risk of being pedantic we will be more explicit. Consider a scalar field $\phi$, together with its conjugate momentum
$\pi$, and a region of space ${\overline P}$ 
at some fixed time $t$ in Minkowski spacetime. By a ``localization procedure" we mean a rationale that 
relates the physical volume $\pb$ to its quantum degrees of freedom $P$ by partitioning the total Hilbert space $\cH$ of the field into $\cH_P \otimes \cH_R$.   If $\p$ is a point in $\pb$  i.e. 
$\p \in \pb$ and $\rr$ is not, i.e. $\rr \in {\rb}$, 
then from the usual commutation relations we clearly have that 
\begin{equation}
[\phi(\p),  \phi(\rr)]\ =\
[\pi(\p),  \pi(\rr)]\ =\
[\phi(\p),  \pi(\rr)]\ =\ 0\, .
\end{equation}
Also linear combinations of $\phi$, $\pi$ and 
their spatial derivatives
commute if they belong to the two separate regions $\pb$ and ${\rb}$. 
In other words, relation \eqref{commute} is satisfied if we take as the algebra of operators
${\cal A}(P)$ the one generated by the local fields in $\pb$. 
We call the corresponding partition the \emph{standard TPS} or the \emph{standard localization scheme}. 

Before introducing the Newton-Wigner localization scheme we first specify the Hamiltonian $H$ 
of the field system. For simplicity we consider a free scalar field $\phi$ of mass $m$:
\begin{equation} \label{hamiltonian}
H\ =\ \int d^3 k \, w_k \, \ad_\kk\, a_\kk,
\end{equation}
where the usual infinite vacuum contribution has been subtracted, $w_k = \sqrt{{\bf k}^2 + m^2}$ 
and operators $a_\kk$ satisfy 
the commutation relation $[a_\kk, a_\kp]=0,\,  [a_\kk, \ad_\kp]=\delta^3(\kk - \kp)$.
The non self-adjoint 
Newton-Wigner fields $a(\x)$ are just defined as the Fourier transform of $a_\kk$:
\begin{equation} \label{newton}
a(\x) = \frac{1}{(2 \pi)^{3/2}} \int d^3 k \, a_\kk\, e^{ i \kk \cdot \x} , \qquad
\ad(\x) = \frac{1}{(2 \pi)^{3/2}} \int d^3 k \, \ad_\kk\, e^{- i \kk \cdot \x} .
\end{equation}
On the other hand, in the definition of the relativistic fields $\phi$, 
the invariant relativistic measure $(2 w_k)^{-1/2}$ appears in the integral, namely:
\begin{equation} \label{field}
\phi(\x) =  \frac{1}{(2 \pi)^{3/2}} \int \frac{d^3 k}{\sqrt{2 w_k}}
\left(a_\kk  e^{i \kk \cdot \x} + \ad_\kk e^{- i \kk \cdot \x}\right)\, .
\end{equation}
Eq. \eqref{newton} can be seen as a Bogoliubov transformation that doesn't mix creators with 
annihilators and therefore doesn't change the particle content of the system. As for any 
Bogoliubov transformation the commutation relations are preserved, i.e. 
$[a(\x), a(\xp)]=0,\,  [a(\x), \ad(\xp)]=\delta^3(\x - \xp)$.
As before, if $\p \in \pb$ and $\rr \in {\rb}$ 
(i.e. $\rr \notin \pb$), we have  
\begin{equation} \label{commu}
[a(\p),  a(\rr)]\,=\
[a(\p),  \ad(\rr)]\, =\ 0\, ,
\end{equation}
so that the subalgebras produced by the Newton Wigner fields also induce a TPS on $\cH$. (Also, for instance, the operators $a_\kk$ induce a TPS, but it goes without saying that such TPS is a ``less localized'' one, being associated with modes of given momentum). 

\section{Some Properties of the two Schemes}

\subsection{The Hamiltonian is Non-Local in NW}

Perhaps the most striking difference between these two schemes is that interactions are 
local in the standard localization scheme but not in the Newton-Wigner one.
The Hamiltonian is in fact a sum of pieces that are local only in the standard TPS:
\begin{equation} \label{hamlocal}
H\, =\, \int d^3 x \, H(\x)
\end{equation}
(here $H(\x)$ is the Hamiltonian density), but not in the Newton-Wigner one.
For instance, in the case of a free scalar field, from eqs. \eqref{hamiltonian} 
and \eqref{newton} we have
\begin{equation} \label{nonloc}
H\, =\, \int d^3x \, d^3y \, K_{\x \y} \, \ad(\x) a(\y).
\end{equation}
The kernel inside the integral is a function of $|\x - \y|$ that 
dies off as $K_{\x \y} \sim e^{- m|\x - \y|}$ 
for $|\x - \y|\gg m^{-1}$ and $K_{\x \y} \propto |\x - \y|^{-4}$ in the massless case.
Non locality is therefore exponentially suppressed at distances larger than the Compton 
wavelength. 
For massless fields these effects are much more serious although, in the more 
realistic case of several -- massive and massless -- 
fields interacting with each other, 
it is not yet clear to us how to extend the NW localization. 

On the Compton wavelength scales one should also expect violations of causality.
By switching to the Heisenberg picture, 
not surprisingly, NW fields do not commute at spacelike separated events, they do commute only 
if they belong to the same hypersurface $t=const$.
As opposed to $\phi(\x,t)$, $a(\x,t)$, as well as $\pi(\x,t)$, 
are clearly not relativistically invariant objects, 
since their definition depends on a foliation of spacetime into
$t = const$ hypersurfaces that has been choosen at the beginning. A covariant extension 
of $a(\x,t)$ has therefore to include the hypersurface as a variable \cite{flem}. 
In other words, $a$ has to be a function not only of $(\x,t)$ but also
of the future-pointing, unit 4-vector $\eta^{\mu}$ that locally  
represents the observer's quadrivelocity.

\subsection{Every Region of Space has its own Fock Structure in NW}

The Hilbert space of our field theory has a Fock structure:
  \begin{equation}
	\cH = {\mathbb C} \oplus \cH_1 \oplus\ldots \oplus \cH_n\oplus\ldots \;,
	\label{fock}
\end{equation}
where $\cH_1$ is the single particle space and the $n$-particles space, $\cH_n$, is given by the symmetric tensor product of $n$ copies of $\cH_1$. We have seen that a localization scheme is determined when a local algebra of operators ${\cal A}(P)$, corresponding to a volume $\pb$, is specified. In ${\cal A}(P)$, one can always find \emph{ladder operators}, that is, operators that take a vector of $\cH_j$ into one of $\cH_{j+1}$. According to the NW scheme, these are just the NW operators $\ad(\p)$ of eq. \eqref{newton}, with $\p \in \pb$, and their superpositions. In the standard formalism, on the other hand,
one can consider the negative energy part of \eqref{field}:
\begin{equation} 
\phi^-(\p) =  \frac{1}{(2 \pi)^{3/2}} \int \frac{d^3 k}{\sqrt{2 w_k}}
\ad_\kk e^{- i \kk \cdot \p}\, 
\end{equation} 
and superpositions. 
By applying the ladder operators of ${\cal A}(P)$ and ${\cal A}(R)$ to the vacuum state, we find two linear 
varieties $P_1$ and $R_1$ in $\cH_1$, representing the one-particle excitations inside and outside 
$\pb$ according to some localization scheme. Accordingly, the single particle space $\cH_1$ 
decomposes into a direct sum,
\begin{equation} \label{direct}
	\cH_1=P_1\oplus R_1	\, .
\end{equation}
The key point here is that $P_1$ and $R_1$ are not necessarily orthogonal. They are in NW because of the 
commutation relations \eqref{commu} but not in the standard localization scheme, since
the two-point function $\bra 0 |\phi(x) \phi(x')|0\ket$ (without T-product!) doesn't vanish outside 
the lightcone. When $P_1$ and $R_1$ are orthogonal, one can make the identification
\begin{equation} \label{identification} 
P_1 \longrightarrow P_1 \otimes |0\ket_R, \qquad R_1 \longrightarrow |0\ket_P \otimes R_1,
\end{equation}
which, rather intuitively, means that a particle well localized inside $\pb$ leaves $\rb$ 
``empty'' and vice versa.
This is not possible if $P_1$ and $R_1$ are not orthogonal because the RHSs of \eqref{identification} 
are orthogonal by construction.

In the NW scheme we can generalize the identification \eqref{identification} and extend it to the whole 
Hilbert space \cite{fabio, anchesergio}.
The $n$-particles space, being a symmetric tensor product of copies of \eqref{direct}, decomposes into
 \begin{equation} \label{npart}
	 \cH_n=\bigoplus_{k=0}^{n}P_k\otimes R_{n-k} \, ,
\end{equation}
where $P_k$, $R_k$ represent symmetric tensor powers of $P_1$ and $R_1$ respectively. Again,  
the intuitive interpretation here is ``if I have $n$ particles they can be all in $\pb$ and leave $\rb$ 
empty, or I can have $n-1$ particles in $\pb$ and one particle in $\rb$, or $n-2$  etc\dots''. 
The entire Fock space $\cH$ decomposes into two Fock spaces $\cH_P$ and $\cH_R$:
 \begin{equation} \label{fockdecomp}
	\cH = \bigoplus_{n=0}^{\infty}\cH_n=\bigoplus_{n=0}^{\infty}\bigoplus_{k=0}^{n}P_k\otimes R_{n-k} 
	= \bigoplus_{n,\,m=0}^{\infty}P_n\otimes R_m \equiv \cH_P\otimes\cH_R\;.
\end{equation} 
This is not true in the standard localization scheme, where the corresponding $\cH_P$ and $\cH_R$ are not, 
independently, Fock spaces.

To summarize, in the NW case $P_1$ and $R_1$ are orthogonal subspaces of $\cH_1$ that correspond precisely to the regions of space of first quantization\footnote{Note that in the first quantization formalism regions of space are subspaces of $\cH$, rather than subsystems!}. Thus, in the NW scheme we can fairly interpret each volume as a subsystem with an internal Fock structure compatible with the global one. 
On the contrary, in the standard scheme the state of a particle localized in $\pb$ is not orthogonal to that of a particle localized in $\rb$; as a consequence, we can still consider each volume as a subsystem, but not as a Fock space: particles are not separately defined in $\cH_P$ and $\cH_R$ 
(see also \cite{rov} on this). This is strictly related to the vacuum being entangled in the standard scheme. 
More on this in Sec. \ref{sec:entropy}.

\subsection{NW allows ``Strictly Localized'' States}

The usual scheme seriously challenges any idea of ``localized state''. It sounds very natural to define
a state $|\psi\ket$ as ``strictly localized'' \cite{strict} outside $\pb$ if for any possible observable 
$A$ in ${\cal A}(P)$, $\bra\psi|A|\psi\ket=\bra 0|A|0\ket$. In other words, if we excite some 
degrees of freedom that are ``strictly localized'' outside $\pb$, 
the state of affairs inside $\pb$ is the same as the one 
of the vacuum, and I have no chance of detecting something different from the vacuum 
inside $\pb$ by using the local operations ${\cal A(P)}$. It turns out
that no state with finite energy has this property in the standard localization scheme. The state 
$|\psi\ket \equiv \phi(\rr)|0\ket$ which is commonly described as ``a particle at position $\rr$''
is in fact different from the vacuum in any region $\pb$ with $\rr \notin \pb$, i.e. 
$\rho_P \equiv {\rm Tr}_R  |\psi\ket \bra \psi| \neq {\rm Tr}_R  |0\ket \bra 0|$.
This property, which can be traced back to Reeh-Schlieder theorem \cite{clift}, is related, once again, 
with the fact that the vacuum is entangled in the standard scheme. On the other hand, low energy 
excitations can be ``strictly local'' in the NW scheme because of the factorization 
\eqref{identification} that leaves $\pb$ empty and in its ``local vacuum'' whenever we excite some degrees
of freedom somewhere else (i.e. in $\rb$).

\section{Entropy}\label{sec:entropy}

Although expressed as the integral of a local density, the energy \eqref{hamlocal} hides  a certain amount of non-extensiveness. By isolating, as before, a region $\pb$ from the rest $\rb$ of Minkowski space, one easily realizes that $H \neq H(\pb) + H(\rb)$. Just adding the inside and outside contribution, $H(\pb) + H(\rb)$, in fact, leaves out of the Hamiltonian the UV-divergent contact term coming from the gradients across the boundary of $\pb$. It is because of such interaction terms that 
the vacuum is entangled in the standard localization scheme: 
its Von Neumann entropy is UV-divergent  and proportional to the boundary of $\pb$ \cite{scaling}. 

Von Neumann entropy is also known  (see e.g. \cite{wehrl}) to be the appropriate generalization of thermodynamical entropy for generic quantum states. 
In the case of conformal field theories the Von Neumann entropy of a region/subsystem has been 
calculated for a thermal state $\rho \propto e^{- \beta H}$ in 
1+1 \cite{cala} and -- 
using insights from AdS/CFT correspondence -- also in higher dimensions \cite{ads}. In such QFT calculations,
in order to recover a thermodynamically sensible result (e.g. $S_{\rm therm} \simeq
V  T^3$ for a massless field in 3 dimensions), 
the divergent contribution of the vacuum has to be systematically subtracted.
Such a subtraction procedure, as noted in \cite{callan}, is problematic because of the non-trivial dependence of the correction on area. Moreover, one can construct, starting from the vacuum, 
quite \emph{ad hoc} states of higher and higher energy which are less and less entangled: 
after the subtraction those states would end up having a negative entropy! 

Clearly, a basic issue to be understood is whether or not such a divergent entropy 
actually accounts for practically measurable correlations, i.e. whether or not it has any operational meaning.
If the procedure described in \cite{reznik} to create EPR pairs from
vacuum entanglement  turned out to be experimentally practicable, this would strongly suggest that the 
standard localization scheme is the correct way to isolate local quantum degrees of freedom.

In this respect, the NW localization scheme can be seen as a sort of UV-regulator. 
If we isolate a region of space according to the NW procedure we find in fact that  
the vacuum is a product state $|0\ket = |0\ket_P \otimes |0\ket_R$ 
and the corresponding Von Neumann entropy is zero.  
In the free field case \eqref{hamiltonian}, if we switch the temperature on, 
the (non normalized) reduced density matrix $\rho_P \propto {\rm Tr}_R\, e^{- \beta H}$
is block diagonal in each Fock subspace of given particle number. The trace of its $n^{\rm th}$ power
nicely rearranges  in an exponential, giving \cite{anchesergio}
\begin{equation} \label{exp}
{\rm Tr}_P \, \rho_P^n \ = \ \exp\left(\sum_{j=1}^{\infty}\frac{1}{j} {\rm Tr}\, K^{j n}\right)\, .
\end{equation}
Here $K$ is the two-point function
\begin{equation}
K(\p_1, \p_2)\, \equiv \frac{_P\bra 0| a(\p_1)\, \rho_P\, \ad(\p_2)|0\ket_P}{_P\bra 0|\, \rho_P \, |0\ket_P},
\end{equation}
where $a(\p)$ and $\ad(\p)$ are the Newton Wigner operators \eqref{newton} and $\p_1$ and $\p_2$ are points 
inside $\pb$. The trace on the RHS of \eqref{exp} is made inside subsystem $P$ 
and limited to one-particle subspace:
\begin{equation}
{\rm Tr}\, K^m \ \equiv \ \int_{\p_1 \dots \p_m \in \pb} dp_1 dp_2 \dots dp_m\, K(\p_1,\p_2)K(\p_2,\p_3) \dots K(\p_m,\p_1).
 \end{equation}
From \eqref{exp} we can then use the trick 
(see e.g. \cite{callan})
\begin{equation} \label{entropy}
S \, \equiv\,  -{\rm Tr}_P (\rho_P \ln \rho_P)\, = \, \left. \left(-\frac{d}{d n} + 1\right)\ln {\rm Tr}_P \, \rho_P^n \right|_{n=1}
\end{equation}
to calculate the Von Neumann entropy $S$. While referring to \cite{anchesergio} for more details, here we 
point out that the entropy \eqref{entropy} -- in the NW localization scheme -- 
is a thermodynamically sensible quantity: it
doesn't have UV divergences, it vanishes at zero temperature and gradually increases to reach the $S \sim V T^3 $ behavior (for a massless field)
in the high temperature $T\gg V^{-1/3}$ limit. At no stage have we found an area dependent contribution. 

\section{Conclusions}

We have considered two different localization schemes i.e. two different ways of 
relating some physical volume $\pb$ to its quantum degrees of freedom $P$ by partitioning 
the total Hilbert space $\cH$ of the field into $\cH_P \otimes \cH_R$.
We stress again that going from one tensor product structure $\cH_P \otimes \cH_R$ 
to another is not like playing with the points
of space across the border of $\pb$, or choosing some different smearing or compact support
function for our definitions. As explained in the introduction, changing TPS is deeper 
than ``playing with the parts'' of a set in the usual intuitive sense: here $\pb$ is a 
subset of $\mathbb{R}^3$, $P$ is a subsystem.

As long as we are concerned only with the internal dynamics of the fields,
all TPSs describe precisely the same state of affairs: we are just considering different -- equally valid --
partitions into subsystems of the field system,
not changing its dynamics (cross sections, decay rates etc\dots). Things may possibly be different when also gravity is 
taken into account. In the standard approach, gravity is included in
the action principle $ S = \int d^4 x\, (R + {\cal L}_{\rm matter})$, solidly  
binding us to the standard localization scheme and to a local (gravity + matter) theory.
Rather adventurously, one may instead stick with the genuine and naive idea that (semiclassical) gravity is really just the geometry of the physical spacetime. Then it would be 
 crucial to understand how the matter degrees of 
freedom feeding into Einstein equations are ``localized'' in the physical spacetime
itself. By incorporating alternative localization schemes, semiclassical gravity inherits from the matter fields a certain amount of non-locality (see eq. \ref{nonloc}), although a consistent formulation of such a non local theory (gravity + matter) has yet to be written and surely calls for a major breakthrough.

\section*{Acknowledgments}
It is a pleasure to thank Michele Arzano, Sergio Cacciatori, Lucien Hardy, Justin Khoury, Matthew Leifer, 
Simone Speziale and Andrew Tolley for exciting discussions and help. F.C. would like to thank Perimeter Institute for hospitality
during the Undergraduate Research Project of summer 2007.
This research was supported by Perimeter Institute for Theoretical Physics.  Research at Perimeter
Institute is supported by the Government of Canada through Industry Canada and by the Province of
Ontario through the Ministry of Research and Innovation.

\end{document}